\documentclass[12pt]{article}
\usepackage{amsmath,euscript,array,amssymb,cite}
\newcommand{\PSbox}[3]{\mbox{\rule{0in}{#3}\includegraphics{#1}\hspace{#2}}
}
\newcommand{\beq}{\begin{eqnarray}}
\newcommand{\eeq}{\end{eqnarray}}
\begin{document}

\baselineskip=18pt

\begin{titlepage}
\begin{center}
         \hfill {\tt hep-th/0204075}\\
       \hfill {CALT-68-2380}\\
\hfill{UPTP-02-02}
\vskip 1.2in

{\LARGE \bf The strength  of small-instanton amplitudes \\ \vskip .1in
      in (deconstructed) gauge theories\\ \vskip .1in
 with compact     extra dimensions  }

\vskip .5in
{\bf
      Erich Poppitz$^a$ and   Yuri Shirman$^b$}

\vskip 0.2in

$^a${\em Department of Physics,
University of Toronto, Toronto, Ontario, M5S1A7, Canada}
\vskip 0.1in

$^b${\em Department of Physics,
California Institute of Technology, Pasadena, CA 91125, USA}
\vskip 0.1in

{\tt
poppitz@physics.utoronto.ca, yuri@theory.caltech.edu}
\end{center}

\vskip .5in
\begin{abstract}
We study  instanton
effects in theories with compact extra dimensions. We perform an
instanton calculation
in a 5d theory on a circle of radius $R$,
with gauge, scalar, and fermion fields in the bulk of the extra
dimension.
We show
that,  depending on the matter  content, instantons of size $\rho \ll
R$   can dominate the amplitude.
Using deconstruction as an ultraviolet definition of the
theory  allows us to show,
in a controlled approximation,  that for a   small number of
bulk fermions, the amplitude for small size instantons  exponentially
grows as $e^{{\cal{O}}(1)R/\rho}$.
\end{abstract}
\end{titlepage}

\newpage

\section{Introduction, summary, and outlook}

\subsection{Introduction}

Theories with extra dimensions may play a role in resolving many issues
in particle physics, from the
hierarchy problem \cite{hierarchy} and supersymmetry breaking
\cite{susybreaking},  to the flavor
problem and proton stability \cite{AS}.  In
many models, the extra dimensions are accessible to Standard Model gauge
fields.
If  gauge fields are allowed into the extra dimensions,
nonperturbative effects seen in the low-energy
4d theory may  substantially differ  from those in ordinary  4d
theories.
In 4d gauge theories,  asymptotic freedom of the gauge
coupling typically suppresses the contributions of
small instantons to instanton amplitudes, ensuring that
the axion potential and the amplitudes of
$\Delta B = \Delta L = 3$ processes  in the Standard Model
are independent of the ultraviolet physics \cite{thooft}.

Higher   dimensional gauge theories, however, are  not asymptotically
free. One expects,
therefore, that the ultraviolet independence of  instanton amplitudes
is not a generic feature of  compactified higher-dimensional theories,
and that
nonperturbative effects may receive ultraviolet  contributions.
That  small-instanton amplitudes are modified is
clear: as  the instanton  becomes much smaller
than the radius of compactification  $R$, a large  number  of
Kaluza-Klein (KK)
modes  begin  to ``feel" its presence. The KK modes'  fluctuations in
the small instanton background can grow as the instanton size $\rho$
decreases,
        modifying the dependence
of the instanton amplitude on $\rho$.  It is this modification that
we   study here, extending
      the work  of \cite{Csaki:2001zx}  on nonperturbative
effects in compactified supersymmetric theories
      to    theories without supersymmetry.

Our main result is that bulk
gauge fields and scalars lead to an exponential,
$e^{{\cal{O}}(1)R/\rho}$,
enhancement of
the small instanton ($\rho \ll R$) amplitude, while bulk fermions tend
to similarly suppress  small instantons; which contribution dominates
is   a detailed
question to which we devote the rest of the paper.  A qualitative
understanding of this
result can be obtained by recalling  't Hooft's original calculation
\cite{thooft} of the  instanton amplitude
$e^{-S_{eff}}$ and the relation   between instanton amplitudes and beta
functions
      \cite{Novikov:uc, Novikov:rd}.
Nonzero modes of bosons in the instanton background contribute
``screening" logarithms
of $\rho$
to the effective action $S_{eff}$.  Screening
logarithms decrease\footnote{This terminology comes from the relation
      $S_{eff}(\rho) = 8 \pi^2/g^2(\rho)$, which holds for the instanton
amplitude in massless QCD. A
``screening" $\ln \rho$
      contribution increases the coupling in the ultraviolet,   decreasing
$S_{eff}$  for small $\rho$, and v.v.
for the ``antiscreening" logarithms. For a review,  see
\cite{Vainshtein:wh,  shifman}.} $S_{eff}$   as $\rho$ is decreased,
hence one expects
bulk bosons to enhance the
small instanton amplitude.  On the other hand, nonzero modes of
       fermions  in the instanton background contribute ``antiscreening"
logarithms that
      increase the action for small $\rho$, suppressing the small 
instanton
amplitude.
The main contribution to the enhancement of the small-$\rho$ amplitude
is
due  to the nonzero modes---the KK fields responsible for the effect
are massive and have no zero modes in the instanton  background.

To make sense of the calculation of instanton effects in a 5d theory,
which is neither renormalizable nor
asymptotically free, we use deconstruction as an ultraviolet definition
of the theory \cite{Arkani-Hamed:2001ca,Cheng:2001vd}.
For   the ``moose" theory in a  5d phase,  we use standard 4d instanton
calculus,
      renormalization,  and regularization of the instanton amplitude.
We believe that our conclusion regarding the growth of the small
instanton
contribution is independent of the
ultraviolet completion of the theory.
The advantage of deconstruction is that it
      allows us to exhibit this growth  in a
controlled,  weak coupling  approximation (at finite $N$),
      while avoiding formal manipulations with the
      infinite KK tower.

\subsection{Summary}

In section \ref{susy}, we begin
by recalling some known exact results on   nonperturbative effects in
compactified 5d supersymmetric theories. In
\cite{Nekrasov:1996cz, Csaki:2001zx} it was shown   that
the operators violating the
anomalous $U(1)_R$ symmetry in the low-energy   theory
are due to 4d instantons wrapped around the
compact dimension. The discussion in section \ref{exactsusy} sets
the framework for our calculation, as
it is precisely the contribution of wrapped 4d instantons
     in  nonsupersymmetric theories  that we
are interested in.

We continue in section \ref{rhosusy} by
       discussing in   detail the
$\rho$ dependence of the instanton amplitude
      in a compactified supersymmetric 5d theory, using a continuum theory
description. We show
      that, because of supersymmetric cancellations, the dependence of the
amplitude
on the instanton size is entirely governed by the 4d beta function. The
instanton
amplitude is
therefore insensitive to the ultraviolet and its strength is completely
determined
by 4d infrared dynamics. This ultraviolet insensitivity of the
instanton amplitude in compactified supersymmetric
theories explains why the  Seiberg-Witten curve of the compactified 5d
theory,  found in
\cite{Nekrasov:1996cz}
      without specifying  details of the ultraviolet completion, agrees
with the
deconstructed curve  \cite{Csaki:2001zx}.

For a self-contained presentation, in section \ref{review} we give a
brief review
of  deconstruction and the  deconstructed semiclassical field
configurations
     whose contributions
to the path integral we will consider.

We then turn, in section \ref{nonsusy}, to the main goal of this paper:
studying the small-instanton dependence
of  instanton amplitudes in nonsupersymmetric compactified 5d
gauge theories.
We begin with a discussion of our
approximation  to massive instanton determinants (the ``step-function"
approximation) in section \ref{stepfunction}.
      We  then use this approximation to calculate  the
      instanton amplitude in the pure gauge theory  in section
\ref{puregauge}. We
use  a  deconstructed ultraviolet definition of the theory to regulate
      and renormalize the amplitude.   We show
that the amplitude grows exponentially, as   $e^{R/\rho}$.
Eqn.~(\ref{seffKK}),  displaying this behavior of $S_{eff}$,  is the
main result of section \ref{puregauge}.

We include bulk scalar and fermion fields in the calculation in
section \ref{bulkmatter}. Within the
same approximation  as in section \ref{puregauge}, we find the
dependence of $S_{eff}$
      on the instanton size for arbitrary bulk matter content.
Eqn.~(\ref{seffgeneral2}) shows
the effect mentioned in the introduction---that bulk
bosons enhance, while bulk fermions  suppress the small-instanton
amplitude.

In section \ref{corrections}, we discuss an important issue:  the
corrections to our
``step-function" approximation to massive instanton determinants and the
effect of these
corrections on the
small-instanton amplitude. Using the small- and
large-mass expansions \cite{Carlitz:yj,Kwon:2000kf} of the
instanton determinant, as well as
      a recently proposed interpolating formula \cite{Kwon:2000kf},
we show that these corrections  do not qualitatively alter our
conclusions.

\subsection{Outlook}

It is important to stress that our
result  for the instanton amplitude holds in more general
backgrounds---Scherk-Schwarz, orbifold,
or   warped  compactifications, such as (deconstructed) slices of
$AdS_5$.
Thus, they
have potential  applications to models with Standard Model
gauge fields propagating in the bulk of the extra dimensions. We comment
on the
strength of instanton amplitudes in models with nonsupersymmetric
      orbifolds of supersymmetric theories
at the end of section \ref{bulkmatter},
      but leave a
detailed investigation of instanton effects in more general backgrounds
for future study.

      If all Standard Model fermions
propagate in the bulk of the extra dimensions, our result indicates that
the small instanton
amplitude is exponentially suppressed---at least in the finite-$N$,
weak-coupling deconstructed case, where our calculation is under control.
We can not make a claim as to the strength of instanton-induced
processes in the continuum limit. This is a strong-coupling limit of the
deconstructed theory, where
the semiclassical approximation is not valid. On the other hand,
instanton induced
interactions are allowed by the symmetries of the theory and one can add
the corresponding  terms,
      suppressed by the 5d cutoff scale, to the low-energy lagrangian. Our
calculation, however, does not
allow to determine their coefficient.

A phenomenologically more interesting
situation arises when the Standard Model fermions are localized
in extra  dimension (so that they do not give rise to KK modes, which
suppress the  small instanton amplitude). In this case our finite-$N$,
weak-coupling calculation  indicates that
small-instanton processes might be
unsuppressed. Furthermore,
since the instanton is not localized in the extra dimension,  it
generates a  nonlocal interaction between fermions
localized at different positions in the extra dimensions;  a more
detailed investigation of
these effects  is under way.

\section{Instantons in   compactified 5d gauge theory:
the supersymmetric case}

\label{susy}

In order to set  the framework for our calculation, introduce  notation,
and discuss its relation to previous work,
it is instructive to recall what we know about nonperturbative effects
in  a compactified 5d supersymmetric Yang-Mills   theory.

\subsection{Exact results in
compactified 5d  supersymmetric theory  and semiclassical calculations}

\label{exactsusy}

Consider a 5d $SU(2)$ theory with minimal supersymmetry, corresponding
to $N=2$  in 4d. The continuum version of this  theory was studied
      in  \cite{Nekrasov:1996cz}, where the Seiberg-Witten curve
was found by  symmetry and consistency  arguments. The curve
determines the
low-energy, two-derivative theory of the
KK zero modes, including all nonperturbative effects.
The continuum and deconstructed version of this theory were studied
recently in
      \cite{Csaki:2001zx}, where a  map between the nonperturbative 
effects
in the two descriptions was constructed. It was shown
that the nonperturbative contributions to the Seiberg-Witten curve of
the compactified 5d theory
arise, in the semiclassical regime, from 4d instantons
wrapping the extra dimension.

It is precisely the effect  of 4d instantons wrapping the compact
dimension
      that we want to study in this paper, in a variety of
non-supersymmetric
compactified
5d theories.
The 4d instantons give the leading contribution
to operators violating classically conserved, but anomalous, quantum
numbers: $B-L$ and  the Peccei-Quinn symmetry in the Standard Model, or
the $U(1)_R$ symmetry of the compactified 5d
``Seiberg-Witten" theory  \cite{Nekrasov:1996cz, Csaki:2001zx}.

The main issue when performing an instanton calculation is  the
dependence of the instanton amplitude on the
instanton size $\rho$ \cite{thooft}. Classically, instantons of any size
are solutions of the Yang-Mills
equations. The instanton size $\rho$ is a collective coordinate, which
one  integrates over in the path integral. The integral over instanton
size $\rho$  classically diverges. However, quantum corrections to the
instanton amplitude
introduce $\rho$ dependence in the action that   helps  to
determine  the typical size of  instantons
contributing to the  amplitude.

The discussion of the previous paragraph holds also in   compactified 5d
Yang-Mills theory, so
long as one considers only the leading two-derivative term in the
action: the 4d instanton solution independent on the compact coordinate
is  also a  solution of the 5d Yang-Mills equations for arbitrary
instanton size.
The 5d Yang-Mills theory, however,   is at best an effective low-energy
theory and contains (generally unknown)   higher-derivative terms
suppressed by inverse powers of the cutoff   $M$.
These terms can, a priori, determine the instanton size already at the
classical level (in the manner the Skyrme term determines the size of the
skyrmion, or similar higher-dimensional terms determine the size of 
instantons
considered as solitonic particles in a 5d theory, see 
\cite{Hill:2001bt}).
However, the higher-derivative terms' contribution to the instanton 
action
is suppressed by inverse powers of $\rho M$, hence for
instantons larger than the 5d short-distance cutoff, $\rho M \gg1$,
their effect on the action is negligible.

We will see, in section \ref{nonsusy}, that the enhancement of the
small instanton amplitude   in nonsupersymmetric theories  occurs  for
instantons of size
      $R \gg \rho \gg M^{-1}$, where the   higher-derivative terms are
unimportant; parametrically, at least, the required   separation of
scales can always be achieved.\footnote{To discuss
the effects of instantons of size $\rho \sim M^{-1}$, or even $\rho \ll
M^{-1}$,  an  ultraviolet definition of the theory is needed. In a
deconstructed framework,
we give a qualitative discussion of  the  effects of  ``really small"
instantons ($\rho \ll M^{-1}$) in section \ref{puregauge}. In  section
\ref{nonsusy},
the cutoff of the 5d theory is
denoted by $v$, while $M$ is  a Pauli-Villars regulator mass; physical
quantities in
the renormalizable deconstructed theory are independent of $M$.}

In a supersymmetric theory, on the other hand, as we show
below (section \ref{rhosusy}), the
$\rho$ dependence of the
effective action in the instanton background is governed by the
4d beta function. In supersymmetric theories, therefore,  the
instanton amplitude shows no ultraviolet (UV) sensitivity.

\subsection{The $\rho$ dependence of the instanton amplitude in
compactified supersymmetric
theories}

\label{rhosusy}

We  are  interested in   the
      semiclassical expansion of the 5d path integral,
formally written in terms of an infinite number of 4d Kaluza-Klein (KK)
fields,
or the deconstructed version thereof (see section \ref{review}), around
a 4d instanton solution.
Semiclassically, the amplitude is proportional to $e^{- S_{class}}$,
where
$S_{class}$ is the  classical action of the instanton in the 5d theory
with coupling $g_5$, compactified on a circle of radius $R$:
\beq
\label{sclass}
S_{class} = 2 \pi R ~ {8 \pi^2  \over g_5^2 }.
\eeq
Recall now that
in the Wilsonean regularization scheme
the matching to the coupling, $g_4$, of the 4d theory,
occurs at the UV cutoff  $M$:\footnote{And not at $1/R$, as one might
naively
guess; matching at $1/R$ is in conflict with holomorphy in $R$ as noted
in \cite{Luty:1999cz}. This also follows from the RGE
(\ref{deconstructedRGE}) in the
deconstructed version of the theory (with
$m_{KK}=2v\rightarrow M$). Note that the matching
at the UV cutoff is a general feature of deconstructed
regularization of higher dimensional theories even in the
absence of supersymmetry.}
\beq
{1\over g_4^2(M)} = {2 \pi R \over g_5^2}~.
\eeq
Therefore,  expressed in terms of the 4d coupling, the classical action
is:
\beq
\label{a1}
S_{class} = {  8 \pi^2\over g_4^2(M) }~,
\eeq
just like in a 4d instanton calculation with $g_4(M)$---the ``bare"
coupling. As usual in instanton calculations, we will take $M$ to
be the Pauli-Villars regulator mass. The classical action is, of course,
independent of the instanton size, which is a collective
coordinate to be integrated over. However,
fluctuations around the instanton depend on  its size $\rho$. Studying
this dependence is our main goal, since it is crucial in
determining  the strength of the instanton amplitude.

Let us look first at the fluctuations of the
4d gauge field (the KK zero mode) in the instanton background.  The
nonzero
modes of the $n=0$ KK states
cancel due to supersymmetry \cite{shifman}.
The only contribution is that of the zero modes, changing $e^{-
S_{class}}$, (\ref{a1}),  to:
\beq
\label{a2}
\exp\left(-{ 8 \pi^2\over g_4^2(M)}\right)  (\rho M)^{B  - F/2}
\eeq
where $B, F$ are the number of bosonic and fermionic zero modes
respectively.
The way (\ref{a2}) appears is  very simple---the Pauli-Villars regulator
determinants have
modes with eigenvalue $M^2$ for bosons (and $M$ for  fermions) whenever
there is a zero mode.
Since one divides by the PV determinant, regulator boson  eigenvalues
appear with +1/2 power
and regulator
fermions---with -1/2 (instead of -1/2 and 1/2, respectively, for physical
fields). Combining
the above one gets the $B - F/2$ factor in (\ref{a2}).

The important point is that in the supersymmetric case the  only
$\rho$ dependence appears through the zero modes.
There is no $\rho$ anywhere else in the whole Green function (except  in
the  measure, of course). This is to be contrasted with ordinary QCD
where the nonzero modes also contribute a $\rho$-dependent factor, as we
will
see  in section \ref{nonsusy}.
Now, exponentiating the prefactor in (\ref{a2}),  $S_{class}$ becomes:
\beq
\label{a3}
S_{eff} = { 8 \pi^2\over g_4^2(M)} - \left(B  - {F\over 2}\right) \ln
\rho M \
\eeq
The size of the instanton is determined \cite{thooft} by adding the
Higgs
action with vev $a$, proportional to $a^2 \rho^2$, and
then extremizing
w.r.t. $\rho$, which gives $\rho^2 \sim b_0/a^2$. Here, $b_0$ is
the first coefficient of the beta function. In supersymmetric theories
$b_0=B - F/2$ precisely because  nonzero modes in the instanton
background
cancel (for the $SU(2)$ $N=2$ supersymmetric
theory
we are considering $B = F = 8$); see \cite{shifman}.

Now we turn to the fluctuations of the KK modes of the gauge
field and partners.
The  spectrum of fluctuations of the heavy modes (with mass $n/R$)
around the instanton  still has an asymmetry between bosons and fermions.
This is easiest to see from the fluctuation equations. For example,  the
KK modes of the gauge field obey the same equation (in the $A_{5}^{n} =
0$ background Lorentz gauge) as the $n=0$ mode, up to the mass term:
\beq
\label{a5}
- D^2(A^0) A^n_\mu + 2   [F^0_{\mu \nu}, A^{n~ \nu}] + {n^2\over R^2}
A^n_\mu = \lambda_b A^n_\mu \, ,
\eeq
where $A^n$ is the KK gauge field and $A_0$, $F^0_{\mu\nu}$---the BPST
background.
If $R$ is infinite, these are the usual 4d fluctuation equations in the
background Lorentz gauge.
Adding the KK mass
only shifts the spectrum upwards by $n/R$;  the fermions still  have
eight chiral eigenfunctions
with eigenvalues $n/R$ and the bosons--- eight bosonic eigenfunctions
with eigenvalues $n/R$.
There is, therefore, still a net
non-cancellation of the lowest eigenvalues in the determinants at each
KK mass level.
Since all eigenvalues larger than $n/R$  come in complete supersymmetry
pairs   the  contributions of all but the lowest eigenvalues  cancel in
the determinants.

Therefore the fluctuation determinant of the KK
gauge fields reduces to the
product of the lowest eigenvalues  (which would
have been the zero modes in the infinite-$R$  limit).
Taking these KK gauge
fields ``zero modes"
      into account, the instanton amplitude in the
supersymmetric $SU(2)$ theory becomes:\footnote{The  product in
     (\ref{a6}) should be taken over infinite KK
tower. However, we are working in an effective 4d theory with a cutoff
$M$, and absorbed the contributions of the KK states with mass $m_n>M$
into the definition of $g_4(M)$. This becomes obvious when the theory
is regulated by deconstruction: as long as cutoff (lattice spacing)
is finite, the KK tower is finite too.}
\beq
\label{a6}
\exp\left( - {8 \pi^2\over g_4^2(M)} +
4 \ln (M \rho)\right) \prod_{n = 1}^{MR} (n R)^{-8}
\left(\sqrt{n^2/R^2 + M^2}\right)^8~.
\eeq
The first term in the pre-exponential product
is the contribution of the would-be zero modes of physical fields and
the second term---that of the regulators.\footnote{In the
non-supersymmetric case, as we will see in the following sections, we
will also advocate that
the pre-exponential factors have to be present, based on the singularity
of the determinants in the $R \rightarrow \infty$ limit
and/or renormalization group invariance with respect to changes of the
UV cutoff $M$.}
The contribution is to the power of 8 and not 4 because at each KK level
there are left and right moving modes, so the number of zero modes of
the non-zero KK modes at each level
is double that of the $n=0$ zero modes.
Eqn.~(\ref{a6}) should be multiplied by integrals over the usual
fermionic and bosonic zero modes. Since our main interest is
      the $\rho$ dependence of the measure,
we will not explicitly display these integrals here and in the following
sections.

It is important to note that the product in (\ref{a6}) is independent
of $\rho$. We can re-exponentiate and obtain:
\beq
\label{a7}
S_{eff} = {8 \pi^2\over g_4^2(M)} - 4 \ln M \rho - 4 \sum_{n =1}^{MR}
\ln \left(1 + {M^2 R^2 \pi^2\over n^2 \pi^2} \right)
= {8 \pi^2\over g_4^2(1/R)}-4\ln \frac{\rho}{R}~.
\eeq
We conclude that the instanton amplitude is proportional to
\beq
\label{a9}
\exp\left( - S_{eff} \right) = \exp\left( - {8 \pi^2 \over g^2_4(1/
R)} + 4 \ln{\rho \over  R }
\right) ~ = ~ \left( \Lambda_4\,  \rho \right)^4~,
\eeq
where $\Lambda_4$ is the 4d strong coupling scale of the theory as
defined by (\ref{a9}).

Eqn.~(\ref{a9}) is the main result of this section. It shows that the
dependence of the
instanton amplitude on the instanton size in the compactified
supersymmetric
5d theory   is entirely controlled by the 4d beta
function. The size of the relevant instantons is then  determined
by the 4d infrared (e.g. Higgs) dynamics of the theory. The
supersymmetric cancellations of
      nonzero modes    are crucial in establishing this result.

\section{A brief review of deconstruction }

\label{review}

In this section we briefly review the idea of
deconstruction---the regularization that we will use  to perform
the instanton calculation in the compactified nonsupersymmetric theory.
    Deconstruction has been proposed in
\cite{Arkani-Hamed:2001ca,Cheng:2001vd} as a convenient,
explicitly gauge invariant regularization scheme for the
study of $D=4+n$ dimensional gauge theories with $n$ compact
dimensions.
In this paper we study 5d theories compactified on
$S^1$ with
radius $R$.
Their deconstructed description can be obtained by
discretizing a single compact dimension. With $N$ lattice
sites the discretized lagrangian of a pure gauge theory with
an $SU(M)$ gauge group has the form:
\beq
\label{lattice}
S=\int\,d^4 x\left(  -\frac{a}{g_5^2} \, \sum_{k=1}^{N} {\rm tr} F^2_{k}+
\frac{a}{g_5^2} \, \sum_{k=1}^{N} {\rm tr} (D_\mu Q_k)^\dagger D^\mu
Q_k \right)~,
\eeq
where $a$ is a lattice spacing, and
the link fields $Q$ are proportional to Wilson lines:
\beq
Q=\frac{1}{a}~e^{iaA_5}~.
\eeq
It is obvious that the action (\ref{lattice}) describes a 4d
product gauge theory with $SU(M)^N$ gauge group and scalar
fields $Q$ in the bifundamental representations. More precisely,
     the product gauge theory description requires that
the lagrangian (\ref{lattice}) be supplemented by a scalar
potential $V(Q)$,    generating non-vanishing vacuum
expectation values (vevs) for the link fields $<Q>=v=1/a$.
The dictionary relating the 5d and 4d parameters is given by:
\beq
\label{dictionary}
a=\frac{1}{v},~~~
2\pi R=\frac{N}{v},~~~
\frac{1}{g_5^2}=\frac{v}{g^2},
\eeq
where $g^2$ is a gauge coupling of an individual gauge group
in product gauge theory. Note that the scalar kinetic term
has a non-canonical normalization, $1/g^2$, which is
especially convenient in a supersymmetric theory.
In fact, one can generalize deconstruction to study
supersymmetric theories in 5d
\cite{Csaki:2001em}. In this case, there is a 4d SYM theory on
each lattice site. In the simplest case, when four supercharges are
preserved
by the latticized theory,  the link fields $Q$ are
chiral superfields.

Expectation values of the link fields break the product gauge group
$SU(M)^N$ to the diagonal $SU(M)$,
giving mass to all but one linear combination of the gauge
multiplets. The eigenvalues of the mass matrix are given by:
\beq
m^2_k= 4 \, v^2 \, {\rm  sin}^2 \frac{\pi k}{N}~,~ k = 0, \ldots N-1.
\eeq
One can easily see,
     using the identifications  from eqn.~(\ref{dictionary}),  that, in
large-$N$ limit, the KK
spectrum of a 5d theory is reproduced.

We now turn to the gauge coupling of the unbroken $SU(M)$
gauge group, which should be identified with the low energy
gauge symmetry of the compactified 5d theory. The matching
of the couplings should be performed at the scale of the
highest KK mode $m_{KK}=2v$ and is given by:
\beq
\frac{1}{g^2_d(m_{KK})}=\frac{N}{g^2(m_{KK})}~.
\eeq
Note that the continuum (large-$N$) limit with  fixed 5d coupling
and radius (which at the same time implies fixed coupling $g_d^2$ in
the low energy theory) requires that:
\beq
\label{strong}
g^2(m_{KK})\sim v\sim N\rightarrow\infty~.
\eeq
Eqn.~(\ref{strong}) implies that the strong coupling scale of each
individual gauge group
in the product gauge theory also goes to infinity along with $N$. Hence,
     in the continuum limit the deconstructed description of the theory is
not weakly coupled, reflecting the non-renormalizability of the
original 5d  model.

On the other hand, the gauge coupling of the diagonal subgroup formally
appears asymptotically free   and its  renormalization group (RG)
evolution is given by:
\beq
\label{deconstructedRGE}
\frac{8\pi^2}{g_d^2(1/R)} &=& N\, \frac{8\pi^2}{g^2(m_{KK})}- b_0\, \ln
m_{KK} R
      + 2 b_0 \, \sum_{k=1}^{\frac{N-1}{2}} \ln \frac{m_k}{m_{KK}}
\nonumber \\
&=& N
      \frac{8\pi^2}{g^2(m_{KK})} - b_0 \, N \, \ln 2 + b_0\,  \ln 2\pi~,
\eeq
where as before
$b_0$ is the one-loop beta function coefficient
in the low energy theory
(e.g., $b_0=4$ in the 5d $SU(2)$ SYM model of
section \ref{susy} and $b_0=7$ in the non-supersymmetric $SU(2)$ pure
gauge
theory of
section \ref{puregauge}), and the factor of two in the sum is due to
the left and right moving modes. The first line of the
RG equation (\ref{deconstructedRGE}) is written, for definiteness, for
odd $N$ (the second line
is true for any $N$).

We now turn to the discussion of non-perturbative physics. We are
interested in   the contributions of instantons in the
compactified 5d theory
to amplitudes in the low-energy effective theory.
In fact, truly localized 5d
instantons are not known. However, as has been explained in
Section \ref{susy}, topologically non-trivial solutions of
the classical equations of motion can be obtained by
``lifting'' 4d instanton solutions to the fifth dimension. Namely, one
should consider field configurations which are independent of the fifth
coordinate:
\beq
\label{instanton}
A_\mu(x_\nu,x_5)=A_\mu^{\mathrm{inst}}(x_\nu),~~~A_5(x_\nu,x_5)=0~.
\eeq
This 5d ``instanton'' is a codimension-4 object, whose world-line wraps
the compact dimension. Clearly, the action is finite so long as the
compactification radius $R$ is finite.

We are looking for the description of such instantons in the
deconstructed theory. The independence of the compact
coordinate in the continuum theory suggests that in the deconstructed
description we should look for field configurations independent of
position in ``theory space" (that is, the gauge fields in all
of the gauge groups should have the same profile). In other words,
we are looking
for a multi-instanton solution with  winding numbers
$(1,1,\ldots,1)$ in the individual gauge groups of the
$SU(2)^N$ theory.\footnote{See \cite{Csaki:1998vv} for an explanation of
this terminology.}
In the low energy effective theory this
field configuration corresponds to a single instanton in the unbroken
gauge group, as   expected from the wrapped 5d instanton picture.
Indeed, it was shown in \cite{Csaki:2001zx} that in the continuum limit
this multi-instanton solution corresponds to the 5d instanton
(\ref{instanton}). On the
other hand, the deconstructed theory has a much richer non-perturbative
physics. In particular there exist ``fractional'' instantons with
winding numbers $(1,0,\ldots,0)$ and so on. Such fractional instantons
are lattice artifacts which do not correspond to physical effects in the
continuum theory.
In \cite{Csaki:2001zx} it
was shown that in supersymmetric theories fractional instanton
contributions cancel  and do not affect  the  low-energy
action (ensuring agreement with the continuum theory result
\cite{Nekrasov:1996cz}
for the low-energy Seiberg-Witten curve).
However, one expects that in non-supersymmetric models fractional
instantons will affect the low energy properties of the
theory. Nevertheless one can use deconstruction to study instanton
amplitudes of the 5d theory if one only includes contributions of
$(1,1,\ldots,1)$ instantons to the path integral.
It is in this sense that we will treat deconstruction as an UV
definition of the 5d theory in the following sections.

\section{Instantons in   compactified 5d gauge theory: the
non-supersymmetric case}

\label{nonsusy}

In this section, we   study the $\rho$ dependence of the instanton
amplitude
in a non-supersymmetric compactified
5d theory.
The main difference from the supersymmetric case
is that nonzero modes  contribute at all KK levels---without
supersymmetry
      no complete cancellation between fermions and bosons can occur.
The results of this section apply to purely non-supersymmetric
models as well as to models where the field content is supersymmetric,
but the spectrum is not, e.g.
Scherk-Schwarz and orbifold models. We assume that the expectation value
of the
Wilson line ${\cal{P}} \exp \int_0^{2 \pi R} A_5 (x,y)   d y$ vanishes
or is small enough to
      not affect the $\rho \ll R$ behavior of the instanton amplitude
(alternatively,
      the Wilson line can be projected out by orbifolding).

We will consider  the  instanton amplitude in the pure-gauge
(deconstructed) compactified 5d
theory in section \ref{puregauge} and include bulk matter  in section
\ref{bulkmatter}.
However, we need to first
      make a detour and briefly discuss
      the evaluation of KK mode determinants   in the instanton 
background,
in particular
the ``step-function" approximation that we will employ in our
calculation. In
section \ref{corrections}, we will give a more detailed discussion of
the corrections
to this approximation and how they affect our results.

\subsection{The ``step-function" approximation to massive instanton
determinants}

\label{stepfunction}

Consider first an instanton in a 4d $SU(2)$ gauge  theory with a
scalar field of mass $m$, which we take (for definiteness) to be a
    complex adjoint   field. We write down
general form of the instanton action as:
\beq
S_{eff} = \frac{8\pi^2}{g^2(M)}- \left( 8- {2\over 3} \right) \ln
M\rho +\Delta S_{eff}(m\rho, M/m)
\,
\eeq
where $M$ is a UV cutoff (a Pauli-Villars mass, $M\gg m$,  $M\gg \rho$),
and $\Delta S_{eff}(m\rho, M/m)$ is the
contribution of the scalar field fluctuations  around the
instanton.
The  antiscreening factor of $-8$  is due to the zero modes of the gauge
field in the instanton background,
while the screening factor of $+2/3$  is from the nonzero modes of the
gauge field  \cite{thooft,Vainshtein:wh}.

To determine   $\Delta S_{eff}(m\rho, M/m)$, we would have to
calculate the determinant of  scalar fluctuations for arbitrary $m\rho$.
This is a technically difficult task---the
method of \cite{thooft}  uses   the conformal invariance of
the massless case and is not  applicable to the massive determinant.
However,
      continuity in $\rho$, renormalization group invariance with
respect to a change in the cutoff $M$,  and the singularities of the
determinants expected in the massless limit,
will help us find an approximate answer.
\begin{enumerate}
\item{
For
$m\rho\ll 1$, we can neglect the mass  and
therefore $\Delta S_{eff}(m\rho, M/m)= {2\over 3} \ln
M\rho$;
the coefficient is the
one appropriate for a complex adjoint scalar
found in \cite{thooft}. An important point is
that $\Delta S_{eff}(m\rho, M/m)$
has a smooth limit as $m \rightarrow 0$ for fixed
$\rho$.
This is because the scalar field does not have zero modes in the
instanton background
even in the massless limit \cite{thooft}. Such a zero mode would cause a
divergence of the instanton amplitude in the $m \rightarrow
0$ limit.  This is to be contrasted with the case of a KK gauge
field, where bosonic zero modes in the instanton background will appear
in the massless limit, causing a singularity in the
instanton amplitude; we will see that in
the case of massive vector fields
$\Delta S_{eff}$ should be modified to account
for the massless limit singularity. There are corrections to the
small-mass expression we use:
a constant piece, calculated by 't Hooft as well as corrections
proportional to positive powers of $m \rho$; these are discussed in
section \ref{corrections}, see
eqn.~(\ref{smallmass}) there.
}
\item{For  $m\rho \gg 1$, the  scalar does not feel the instanton, and
therefore $\Delta S_{eff}(m\rho, M/m) = const (M)$,
(independent of $\rho$) where the dependence on $M$ is such as to make
the action renormalization group invariant with respect to changes of
$M$. Corrections suppressed by $1/(m \rho)$ are certainly
present, but are small when $m \rho \gg 1$, see eqn.~(\ref{largemass})
of section \ref{corrections}.}
\item{Finally, there is another piece of information: quantum
mechanically, the action depends on the instanton
size $\rho$. It is reasonable to require that the
action be a continuous function of $\rho$.}
\end{enumerate}
Combining the above arguments, we are led to the following
expression for $\Delta S_{eff}(m\rho, M/m)$:
\beq
\label{fmrho}
\Delta S_{eff}(m\rho,M/m)=
\frac{2}{3}\ln M\rho + \frac{2}{3}g(m\rho)\, ,
\eeq
where $g(m\rho)$ is approximated by:
\beq
\label{stepapprox}
g(x) = \left\{ \begin{array}{cc} 0, & x < 1 \cr
- \ln x , &   x > 1 \end{array}~. \right.
\eeq
We will further call eqn.
(\ref{stepapprox})   the ``step-function" approximation
(despite the obvious continuity of $\Delta S_{eff}(m\rho, M/m)$ in
$\rho$)  to the determinant.
This is very similar to what one usually does with
beta-functions: the states with masses near cutoff do not of
course decouple discontinuously as the energy scales are
integrated out; yet we discontinuously change the beta
function across the threshold and then require that the gauge
coupling itself be continuous.

The limits considered above will be important for our study of
instanton effects in the (deconstructed) 5d theory.  We will show, in
section \ref{corrections}, that the ``threshold" corrections to the step
function approximation
(\ref{stepapprox}) do not qualitatively
modify our conclusions.

\subsection{Deconstructing  the instanton amplitude in a compactified
pure gauge theory}

\label{puregauge}

With the   discussion of the previous section  in mind, we are   ready
to turn to the
problem at hand---the strength of instanton effects in
the deconstructed   non-supersymmetric pure gauge 5d theory
       \cite{Arkani-Hamed:2001ca, Cheng:2001vd}.
We will organize our discussion  according to the ratio of the instanton
size $\rho$ to the   ``UV cutoff" of the  5d
theory\footnote{The 5d ``UV cutoff" $v$
is not to be confused with the true cutoff of the 4d
deconstructed description, the Pauli-Villars mass  $M$,
which
we take to be larger than $ v$.}  $v$. The scale $v$ is  where the
$SU(2)^N$ theory is broken down to the diagonal $SU(2)_D$ gauge group
(which
is to be thought as the ``5d" gauge group).
We will first consider
the contribution of instantons in the far UV region, beyond the cutoff of
the ``5d" theory, i.e. $1/v \gg \rho$. We will then
increase the size of the instanton and will consider the instanton
amplitude
for  intermediate values of  $\rho$, $2 \pi R=N/v \gg \rho > 1/v$. This
will
turn out to be   the most interesting,   5d,   regime. Finally,
the contribution of large instantons $\rho > R = N/(2 \pi v)$ is
governed by the
four dimensional
beta function and is considered in \cite{thooft}.

We begin with ``really small" instantons, $\rho \ll v^{-1}$, which
effectively means $v=0$, i.e.   with the moose theory in a ``non-5d"
phase.
In this case, there are $N$ independent instantons of the type
$(1,0,0\ldots)$, $(0,1,0, \ldots)$, etc.
Our primary interest, however, are instantons,
which
appear as
solutions of the 5d Yang-Mills equations of motion, independent of the
compactified coordinate.
In other words, we are interested in
      instantons of the  diagonal gauge group in the moose theory.
These are the   $(1,1,1,1....,1)$-instantons---composite objects made
out of $N$
instantons, one in each $SU(2)$ gauge group.
The $N$-instanton one-loop effective action  for $v = 0$
is then (denoting the sizes of various instantons by $\rho_i$):
\beq
\label{vzero}
\prod\limits_{i = 1}^N \exp\left(-\frac{8\pi^2}{g^2(M)}+\left(8-{2\over
3}\right)\ln M\rho_i  -{1\over 3} \ln M\rho_i \right)
= \exp\left(-\frac{8\pi^2 N}{g^2 (M)}+7 \sum\limits_{i=1}^N \ln M\rho_i
\right) \,
\eeq
where in the first equation $8-2/3$ is the contribution of
zero and non-zero modes of the gauge fields, and $-1/3$ is
the contribution of the bifundamental scalars (they are
real) \cite{thooft}.
The instantons are independent, so one should integrate over
$N$ instanton sizes, centers, and orientations, which makes the
$\rho$-integral
extremely UV convergent and infrared (IR) divergent.
As $v$, breaking $SU(2)^N$ to $SU(2)_D$, is turned on, there is really
only one true zero
mode;  however, for small $v$ one still has $N$ constrained
instantons, and the expression (\ref{vzero}) is a good
approximation.

However,  when $v \ne 0$, separating instanton centers and/or
varying instanton sizes $\rho_i$ independently increases the
action---the mass term $  v^2 \sum_i (A_i^\mu - A_{i - 1}^\mu)^2$
increases the action whenever $A_i^\mu \ne A_{i-1}^\mu$.
Therefore, one can   integrate over
$N-1$ instanton sizes and positions (or rather their
variations from the ``average'') and end up with the action of
a single $(1,1,\ldots,1)$ instanton, so that one only
integrates over a single $\rho$. From a 5d perspective, this is
equivalent to considering the contribution of an instanton
of the KK zero modes of the 5d gauge field.

The integrand for the diagonal  instanton of size $\rho \ll 1/v$ is, in
our step-function approximation
(\ref{fmrho}):
\beq
\label{vbigger}
\exp\left(-{8\pi^2 N \over g^2(M)}+8\ln M\rho -
\left(1+2 ~{N-1 \over 2}\right)\ln M\rho \right) \times
\prod\limits_{k=1}^{N-1\over 2}
\left({\sqrt{m_k^2+M^2}\over m_k}\right)^{16}~.
\eeq
The various factors in (\ref{vbigger}) arise as follows:
\begin{enumerate}
\item{The antiscreening logarithm $8 \log M \rho$ reflects the
contribution of the true zero modes of
the diagonal unbroken gauge group.}
\item{
The screening logarithm comes, as usual,
from the contribution of non-zero modes of both massless and
KK states---recall that $\rho\ll 1/v$,  therefore all states are lighter
than $1/\rho$ and
feel the presence of the instanton. More precisely,
the screening contribution from the massless states---the diagonal gauge
bosons' nonzero modes as well
as the massless  real adjoint scalars' nonzero modes---gives a total
factor of $-1=-2/3-1/3$.
The $N-1 \over  2$ (for simplicity $N$ is taken odd here)
doubly degenerate (left and right movers) KK levels contribute  each the
same factor of $-1 (= -2/3 - 1/3)$,
       resulting in the second term in the screening logarithm.}
\item{
The pre-exponential factor in (\ref{vbigger}) is due to the gauge field
KK states. It
reflects the fact that the $m_k \rightarrow 0$ limit
of the KK gauge field determinant is singular because of the $16$  zero
eigenvalues appearing at each
KK level; recall that the KK mass levels are doubly degenerate ($16 = 2
\times 8$).
The additive factor of $m_k^2$ in the numerator should, of course,  be
dropped since the UV cutoff $M$ is above any of the KK masses. The
pre-exponential term in (\ref{vbigger})  was written
in a manner underscoring the analogy with the supersymmetric case, where
the lowest eigenvalues are the only ones contributing. Similarly, here
the pre-exponential term can also be thought of as the
contribution of the lowest eigenvalue---the one vanishing in the
massless limit---of
the KK gauge field determinant, ensuring the correct singular behavior
in the $m_k \rightarrow 0$ limit. Further corrections
to the amplitude (\ref{vbigger}) should vanish in the massless limit and
are proportional to positive powers of $m_k \rho$  (see section
\ref{corrections}).}
\item{Finally, it is easily verified  that (\ref{vbigger}) is RG
invariant with respect to change of UV cutoff $M$; the relevant RG
equation is
the one for the individual $SU(2)_i$ gauge
groups, since $M$ is above any mass thresholds.}
\end{enumerate}

Let us now increase size of the instanton. As $1/\rho$
crosses any given threshold $m_k$, the corresponding states
decouple from the instanton, and the contribution of their
non-zero modes simply turns from $-2 \ln M \rho$ into $-2 \ln
(M/m_k)$, as in eqn.~(\ref{fmrho})
in the step function
approximation (\ref{stepapprox}).
Therefore, denoting by $K(\rho)$ the
number of KK levels lighter than  $1/\rho$,
the instanton amplitude (\ref{vbigger}) turns into:
\beq
\label{vsmaller}
\exp\left(-{8\pi^2 N \over g^2(M)}+8\ln M\rho -
(1+2K(\rho))\ln M\rho  \right)  \times   \prod\limits_{k=1}^{K(\rho)}
\left({M \over m_k}\right)^{16}~
\prod\limits_{k=K(\rho)+1}^{N-1\over 2}
       \left({M\over m_k}\right)^{14}\,.
\eeq
The pre-exponential factor for the gauge fields heavier than $1/\rho$
is required by RG invariance of the amplitude with respect to changes
of  $M$; since
       states with mass greater than $1/\rho$ do not feel the instanton,
the
prefactor does not depend on $\rho$ (up to
the already mentioned
       corrections in the action, suppressed by  inverse powers of $\rho
m_{k > K(\rho)} $ or positive powers of $\rho m_{k < K(\rho)}$).

We then exponentiate the prefactor in (\ref{vsmaller}), and after some
simple algebraic manipulations find the
instanton amplitude $e^{- S_{eff}}$, with $S_{eff}$
given by:\footnote{From now on, unless otherwise noted, $g$ stands for
the
        coupling of the diagonal group.}
\beq
\label{seffective}
S_{eff} = {8\pi^2 \over g^2(1/R)} - 7 \ln {\rho\over R} +
2\sum\limits_{k=1}^{K(\rho)} \ln m_k \rho \,.
\eeq
To arrive at (\ref{seffective}), we used RG invariance with respect to
the cutoff $M$ to replace
$M$ with $m_{KK} = 2 v$ in (\ref{vsmaller}) and then applied the
deconstructed theory RG equation (i.e., eqn.~(\ref{deconstructedRGE})
written
for the theory under consideration):
\beq
\label{RGeqn}
{8 \pi^2 \over g^2(m_{KK})} = {8 \pi^2 \over g^2(1/R)} + 7 \ln m_{KK}
R - 14 \sum\limits_{j = 1}^{N-1\over 2} \ln {m_j \over m_{KK}} = {8
\pi^2 \over g^2(1/R)} + 7 N \ln 2 - 7 \ln 2 \pi~ ,
\eeq
       to express $g(m_{KK})$ in terms of $g({1\over R} = {2 \pi v\over
N})$.
The latter, low-energy coupling is the one to be kept fixed in
physical  applications.

Now let us consider $S_{eff}$ of eqn.~(\ref{seffective}) in two cases:
either
by taking the usual, non-deconstructed,  expressions for KK
masses or by using the deconstructed formula.
When $K(\rho)$ is (moderately, see below) large we can also  approximate
$K(\rho) = \left[ R/\rho \right]$.

We begin by taking $m_k=k/R$---a case which is simple to analyze
analytically.
The action (\ref{seffective}) becomes:
\beq
       S_{eff} = {8\pi^2 \over g^2(1/R)}- 7 \ln {\rho\over R} + 2 K(\rho)
       \ln{\rho\over R} + 2  \ln K(\rho)!~.
\eeq
Substituting  $K(\rho) =[R/\rho]$ and using  Stirling's formula
to perform the sum, we obtain for the action:
\beq
\label{seffKK}
S_{eff} (R \gg \rho \gg 1/v)&=& {8\pi^2 \over g^2(1/R)} - 6
\ln{\rho\over R} +
2 {R\over \rho} \ln {\rho\over R}
+ 2 {R\over \rho} \ln {R\over \rho} - 2 {R\over \rho} \nonumber  \\
&=&  {8\pi^2 \over g^2(1/R)}  - 2 {R\over \rho}- 6 \ln{\rho\over R} ~.
\eeq
It is obvious that the log-enhanced terms cancel (except for
low-energy contributions coming from physics below
compactification scale).  Most importantly, we note that
the term linear in $1/\rho$  decreases
the action for small $\rho$,  enhancing the small instanton contribution.

We should stress  that eqn.~(\ref{seffKK}) is a valid approximation to
the deconstructed
expression provided: {\it i.)} $1/\rho$ is sufficiently below the top
of the KK tower so that the expressions
for the KK masses are approximately correct and {\it ii.)} $R/\rho \ge
O(10)$, so that there is a large number
of KK modes lighter than $1/\rho$ that feel the instanton.
The conclusion we can draw from (\ref{seffKK}),  is that
for instanton size   much smaller than $R$ the effective action
decreases as
$\rho$ becomes smaller. Thus, the instanton amplitude  grows in the
UV. Eqn.~(\ref{seffKK})
demonstrates the UV sensitivity of instanton induced  amplitudes
in extra dimensional gauge theories.

Earlier in this section, in
      our qualitative discussion of deconstructed instanton amplitudes, we
argued   that
for $\rho \ll 1/v$ the  action again becomes an increasing function of
$1/\rho$ (more precisely, of the $N$ independent sizes $\rho_i$ of the
$N$ instantons
that the diagonal instanton  ``dissociates" into for $\rho \ll 1/v$).
Together with our analysis of the
KK case of the previous paragraph, this  allows us to
conclude that there is an  ``extra-dimensional"  saddle point
of the $\rho$ integral,
presumably of order the inverse cutoff scale.

In order to study the effects of instantons with sizes near the inverse
cutoff of the deconstructed theory
        in more detail, we will  consider the
       more appropriate  deconstructed expressions for the KK masses:
$m_k = 2 v \sin { \pi k \over N}$, $k = 0,...,{N - 1 \over 2}$.
To study the $\rho$ dependence for smaller values of $\rho$ we proceed
to evaluate eqn.~(\ref{seffective})
numerically. We find that the $\rho$-dependent part of $S_{eff}$ becomes
negative at about $R/\rho \sim 10$,
for all values\footnote{Note that   large values of $N$ are required in
order to observe this effect: the validity of (\ref{seffKK})
already requires
$\rho \gg 1/2v$, i.e. $N/\pi \gg R/\rho > O(10)$; for the theory at
hand, numerically the effect appears at $N \sim 60$.}  of $N$,
consistent with (and implied by!)  the KK analysis above.  Further
decreasing the value of $\rho$, the action decreases to further negative
values (see Figure 1),
somewhat faster than  the KK approximation (\ref{seffKK}) would suggest.

      \begin{figure}
\PSbox{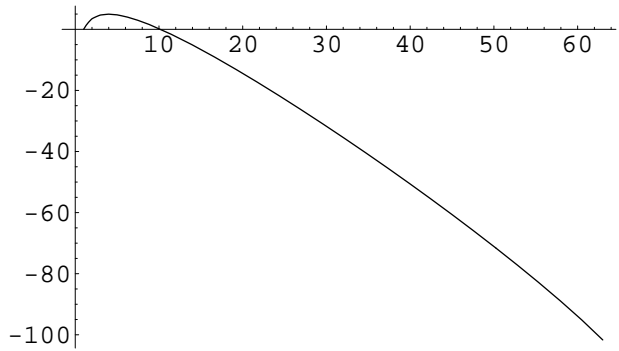 hoffset=30  voffset=0}{8cm}{5cm}
\caption{The $\rho$-dependent part of $S_{eff}$,  (\ref{seffective}),
in the deconstructed ($N = 200$)
5d $SU(2)$ pure gauge theory: $S_{eff} - 8 \pi^2/g^2(1/R)$ for $1
<R/\rho < N/\pi$.}
\label{fig:transform}
\end{figure}

To summarize, our calculation
indicates that instanton amplitudes in extra dimensional gauge theories
grow with  decreasing  instanton size. The strength
of the instanton-induced interaction is thus decided by
the UV completion of the theory. Using deconstruction
as the UV completion of the 5d theory  requires a strong coupling
analysis in the large-$N$ continuum limit. Since the instanton vertex
is allowed by the  symmetries of the theory, one would expect it
to appear in the low energy theory  suppressed by inverse powers of $v$,
the cutoff of the 5d theory.
For a field theory in the strong coupling regime we do not, generally,
expect higher order corrections (even though they are non-calculable) to
conspire to make some operators
smaller than their ``natural" (as determined by dimensional analysis)
value.

Similarly, in more general UV completions of the 5d theory
(such as 5d compactifications of  string theory) one
expects that short distance degrees of freedom may
contribute to amplitudes involving extremely small instantons
($\rho<g_5^2$). However, generically short distance physics
will not suppress\footnote{Theories with UV/IR
correspondence may give a counter example; yet such vacua
are non-generic.} the contribution of the low-energy degrees of
freedom to the instantons of the size $\rho>g_5^2$.

Finally, we note that the results of this section are not qualitatively
changed by orbifolding
the 5d pure gauge theory (e.g. projecting out the zero mode of $A_5$ and
taking only half the KK spectrum),
even though the precise coefficients in $S_{eff}$ do (trivially) change.
In the next section, we consider
the effect matter fields have on the instanton amplitude.

\subsection{Bulk matter and  the instanton amplitude}

\label{bulkmatter}

It is easy to generalize the above calculation to an arbitrary matter
content in the bulk of the
extra dimension. More precisely, we will consider a theory with $N_F^a$
adjoint 5d fermions (i.e.
two 4d adjoint Weyl fermion KK zero modes), $N_F^f$ fundamental 5d
fermions (2 $N_F$ fundamental
Weyl 4d fermion KK zero modes),  $N_S^a$ real adjoint scalars, and
$N_S^f$ complex fundamental
scalars.

Eqn.~(\ref{vsmaller}) for the instanton amplitude
for $R >  \rho > 1/2v$ is trivially generalized
to include the above matter content:
\beq
\label{generalbulkamplitude}
\exp\left[- { 8 \pi^2  \over g^2(M) } + \left( 7 - {8\over 3} N_F^a -
{1\over 3} N_S^a - {2\over 3} N_F^f - {1\over 6} N_S^f  \right) \ln M
\rho \right. + \nonumber \\
\left.  \left(  - 2 + {8 \over 3} N_F^a - {2\over 3} N_S^a + {2 \over 3}
N_F^f - {1\over 3} N_S^f \right)  K(\rho) \ln M \rho \right] \\
\times \prod\limits_{k = 1}^{K(\rho)} \left( {M \over m_k}\right)^{16 -
8 N_F^a - 2 N_F^f} \times \prod\limits_{k = K(\rho) + 1}^{\rm max} \left
( {M \over m_k}\right)^{14 - {16 \over 3} N_F^a - {4 \over 3} N_F^f - {2
\over 3} N_S^a - {1\over 3} N_S^f }  \nonumber~.
\eeq
For simplicity, we assumed that all fields have the same KK spectra
given by $m_k$, $k \ge 1$.
Generalizing (\ref{generalbulkamplitude}) for
nonequal  mass spectra, occurring, e.g., in Scherk-Schwarz/orbifold type
symmetry breaking is easy: one simply has to
separate the preexponential terms corresponding to fields with different
spectra and, after exponentiating,
perform the sums separately.

The amplitude (\ref{generalbulkamplitude}) was derived  as in the pure
gauge
case. The first product in the preexponential reflects the additional
fact that in the massless limit the fermion determinants vanish, while
the second product is as required
by RG invariance. The coefficients for the contributions of the nonzero
modes to the instanton amplitude
for massless scalars and fermions in the adjoint and fundamental
representations
      are taken from  \cite{thooft}; as before,  we are using the
step-function approximation (\ref{fmrho}) for the massive determinants.
For completeness, we give the
contributions of  the zero   and  nonzero modes of  the massless fields
of interest in the table below; we
show the coefficient $c$,  in terms of which the contribution   of the
determinant of the corresponding field  to the
instanton amplitude is $\exp[c \ln M \rho]$:
\begin{center}
\begin{tabular}[h]{l|c|c}
field, representation & zero modes & nonzero modes \\
\hline
gauge, adjoint & 8 &  -$2/3$\\
scalar, real adjoint& 0& -$1/3$\\
scalar, fundamental& 0 &-$1/6$ \\ fermion, Weyl adjoint & -2  & $2/3$\\
fermion, Weyl fundamental& -$1/2$&$1/6$\\
      \end{tabular}
\end{center}
     From the  table,
the amplitude (\ref{generalbulkamplitude}) is easily recovered
      using the considerations of the previous section.

Writing the amplitude as $e^{- S_{eff}}$ and
exponentiating the prefactors, we obtain for the effective action:
\beq
\label{seffgeneral}
S_{eff} &=& {8 \pi^2 \over g^2(M)} +
       \left(- 7 + {8\over 3} N_F^a + {1 \over 3} N_S^a  +{2\over 3}
N_F^f +
{1 \over 6} N_S^f \right) \ln M R  \nonumber  \\
&+&   \left( -14 + {16 \over 3} N_F^a + {2 \over 3} N_S^a + {4 \over 3}
N_F^f + {1 \over 3} N_S^f \right) \sum\limits_{k =1 }^{\rm max} \ln
{M\over m_k}
\\
&+& \left( - 7 + {8 \over 3} N_F^a + {2 \over 3} N_F^f  + {1 \over 3}
N_S^a +{1 \over 6} N_S^f \right) \ln {\rho \over R} \nonumber  \\
      &+& \left( 2  - {8\over 3} N_F^a -{2\over 3} N_F^f + {2\over 3}
N_S^a +
{1\over 3} N_S^f \right) \sum\limits_{k = 1}^{K(\rho)} \ln m_k \rho~.
\nonumber
\eeq
Let us now make some observations on the form of $S_{eff}$. We first
      note that eqn.~(\ref{seffgeneral}) can be used to reproduce the
$\rho$-dependence
of the  instanton amplitude in the supersymmetric case; agreement with
      supersymmetry provides a useful check on our formulae.
A supersymmetric 5d $SU(2)$ theory with $n_a$ adjoint  and $n_f$
fundamental
matter hypermultiplets
corresponds to taking $N_F^a = 1 + n_a$, $N_S^a = 1 + 4 n_a$,  and
$N_S^f = 2 N_F^f = 2 n_f$.
In the supersymmetric case, $S_{eff}$ drastically simplifies---all
$\rho$ dependence is governed by the 4d beta function coefficient, equal
to $- 4 + 4 n_a + n_f$, as
in the pure 5d SYM case we started with:
\beq
\label{seffsusy}
S_{eff}^{SUSY} =  {8 \pi^2 \over g^2(M)} + \left(-4 + 4 n_a + n_f
\right) \left[ \ln M R ~+~
\sum\limits_{k =1 }^{\rm max} \ln {M^2\over m_k^2} ~+~ \ln {\rho \over
R} \right] ~, \eeq
the contribution of the nonzero modes having   cancelled out as in
(\ref{a9}).
The role
of the $\rho$-independent terms in (\ref{seffsusy}) is to change the
argument of
the coupling $M \rightarrow 1/R$, as in (\ref{a9}),  (\ref{seffective}).

We can write eqn.~(\ref{seffgeneral}) in terms of the low-energy 4d
coupling $g(1/R)$ as:
\beq
\label{seffgeneral2}
      S_{eff}  &=&  {8 \pi^2 \over g^2(1/R)}  +
      \left(  - 7 + {8 \over 3} N_F^a + {2 \over 3} N_F^f  + {1 \over 3}
N_S^a +{1 \over 6} N_S^f \right) \ln {\rho \over R} \\
       &+&  \left( 2  - {8\over 3} N_F^a -{2\over 3} N_F^f + {2\over 3}
N_S^a + {1\over 3} N_S^f \right) \sum\limits_{k = 1}^{K(\rho)} \ln m_k
\rho~.\nonumber
\eeq
Eqn.~(\ref{seffgeneral2}) is the main result of this section. It allows
us to study the $\rho$ dependence of the instanton amplitude. As in the
pure gauge case,
the  large-$R/\rho$ behavior  is controlled by
      the last  term, since it is the one   leading to a linear dependence
of the action on $R/\rho$. We can see the behavior at small $\rho$
      already in the KK approximation: if the coefficient of the last term
is
positive, the effective action is an decreasing function for sufficiently
large $R/\rho$, as in the pure gauge case.

The general effect of bulk matter fields is  clear from
(\ref{seffgeneral2}): bulk fermions tend
to suppress the contributions of  small instantons, while bulk bosons
tend to enhance them.
Precisely how the balance in (dis-)favor of small instantons
       is achieved depends on the matter content of the theory.
      For example, for a 5d $SU(2)$ YM theory with $N_F$ fundamental
fermions, equivalent to $N_F$ fundamental flavors
in 4d, we find from eqn.~(\ref{seffgeneral}) that if $N_F < 3$, the
instanton amplitude grows for small $\rho$, while for $N_F \ge 3$,
the amplitude is dominated by the 4d saddle point.

As mentioned in the beginning of this section, eqn.~(\ref{seffgeneral2})
is also applicable for
spectra different from the usual KK spectra, e.g. for orbifold
compactifications where different
fields have different KK expansions. For a 4d theory obtained as an
orbifold of a 5d theory, one can draw
similar conclusions---that bulk bosons enhance the small instanton
contribution, while bulk fermions suppress it.

If one is considering an orbifold of a  supersymmetric theory, one
expects additional significant
cancellations in   (\ref{seffgeneral}). To see this, note that  in the
supersymmetric limit the
term responsible for the growth of the amplitude
for small $\rho$ (the last term in $S_{eff}$) vanishes. If the
supersymmetric theory is orbifolded, one expects the cancellations
between bosons and fermions to still persist, leaving at most a
logarithmic dependence on $\rho$. To this end, consider the contribution
to the instanton amplitude of
      a fundamental matter hypermultiplet,
      where, in the KK approximation,  the scalars have mass $m_k^s = (k +
\theta)/R$ and the fermions---$m_k^f = k/R$, so that
only the fermions have  KK zero modes. Clearly, then, the contribution of
the hypermultiplet
to the last term in (\ref{seffgeneral}) becomes, dropping  possible
$\sim \ln\rho$ contributions arising
from the ``edges" of the spectrum:
\beq
\label{hypercontribution}
{2 \over 3} \sum\limits_{k = 1}^{[R/\rho]}  \ln{ k + \theta \over k }
\simeq {2 \over 3}  \theta ~\ln {R\over \rho}~.
        \eeq
Similar cancellations will occur in   the other orbifolded
supermultiplets, resulting at
most in a logarithmic growth of the amplitude as a function of
instanton size.

Using (\ref{seffgeneral2}), one can also study instanton effects   in
even
more general backgrounds where
gauge fields propagate, such as a slice of  (deconstructed) $AdS_5$. A
more detailed investigation of the implications of (\ref{seffgeneral2})
in different models will
be given elsewhere.

\section{Corrections to the step-function approximation}

\label{corrections}

In this section, we will consider the role of the corrections to the
step-function approximation (\ref{stepapprox})
we used throughout the previous sections. We will show that they do not
qualitatively alter our
conclusions.

Consider first  a bulk scalar field in the fundamental representation,
whose nonzero KK modes
give  the following
contribution\footnote{Note that
we are taking twice
the contribution from the table, because
at each mass level there are two scalars.}
to $S_{eff}$
(as usual, the amplitude
is $\sim e^{-  S_{eff}}$):
\beq
\label{scalar}
\Delta S_{eff}=\frac{1}{3}\sum_{k=1}^{max}\left(\ln M\rho +
      g(m_k\rho)\right) \, .
\eeq
In the step function approximation of eqn.~(\ref{stepapprox}) this
gives rise to the contribution shown in (\ref{seffgeneral}).

Let us now subtract the $\rho$-independent terms (representing the
contribution of the scalar to the
     change of the scale  $M\rightarrow 1/R$  of the gauge coupling) from
the scalar
contribution to the effective action $\Delta S_{eff}$ (\ref{scalar}):
\beq
\label{scalar1}
\Delta
S^\prime_{eff} \equiv \Delta
S_{eff}-\frac{1}{3}\sum_{k=1}^{max}\ln\frac{M}{m_k}=
\frac{1}{3}\sum_{k=1}^{max}\left(g(m_k\rho)+\ln m_k\rho\right)
\,.
\eeq
     $\Delta S^\prime_{eff}$ from eqn.~(\ref{scalar1}) gives the
$\rho$-dependent
part of the contribution of the KK tower
to the instanton action with the gauge coupling evaluated at $1/R$.
    If we substitute (\ref{stepapprox}),
$g(x) =   - \theta(x -  1) \ln x $, into (\ref{scalar1}), we arrive at
the
effective action in the
      step-function approximation: only the light fields   contribute to
$\Delta S_{eff}^\prime$, giving a large negative
($m_k \rho < 1$) contribution to $S_{eff}$ in (\ref{seffgeneral2}).

Here, we are interested in
the corrections to the step-function approximation
(\ref{stepapprox}) to 
$g(m \rho)$ and
     the  effect of the corrections  on the strength of the amplitude.
The small-mass expansion,\footnote{We multiplied $g$ by
     the factor  $1/6$ in
(\ref{smallmass}), (\ref{largemass}) in
order to simplify   comparison with
\cite{Carlitz:yj,Kwon:2000kf},
where the determinant
of a single fundamental scalar, contributing $g/6$, was considered.}
valid for $x \leq  .5$,
of the determinant  is, as shown in \cite{Carlitz:yj,Kwon:2000kf}:
\beq
\label{smallmass}
{1\over 6}~ g(x) = c -  {1\over 2} ~x^2 \left( \ln {1 \over x } + d
\right) + {\cal{O} }(x^4) ~,
\eeq
where $c = 0.145873$, $d = 0.005797$. The large-mass
expansion,  $x \geq 1.8$, computed to order $x^{-8}$ in
\cite{Kwon:2000kf}, is:
\beq
\label{largemass}
{1\over 6}~  g(x) = -{1\over 6} \ln x - {1 \over 75~  x^2} - {17 \over
735 ~x^4} + \ldots
\eeq
Note that, quantitatively, the constant $c$ in the small-mass expansion
is the largest error in the approximation (\ref{stepapprox}):
even if we define the step-function approximation
so that $g(x)=6c$ for all $0<x<1$,
the conclusion that the
scalar field fluctuations enhance the amplitude is not altered (since
$c < 1/6$). It should be noted, however,  that
such a modification of $g(x)$
grossly overestimates the error of the
step-function approximation,
since, as the small $x$
expansion (\ref{smallmass}) shows,
      $g(x)$ is a decreasing function of $x$.

We can be more precise in our estimate of the error if we use the
interpolating function for $g(x)$,
proposed  in
      \cite{Kwon:2000kf} under the plausible
assumption
that the determinant is a smooth function of $x$,   monotonically
connecting
the large- and small-mass limits:
\beq
\label{interpolating}
{1 \over 6} ~g(x)_{interp.} = \frac{0.145873 - 0.443416\,x^2 -
\frac{x^4}{5} + \frac{\ln x}{6}}
        {1 - 3\,x^2 + 20\,x^4 + 15\,x^6} - \frac{\ln x}{6}~,
\eeq
which has the correct expansions (\ref{smallmass}), (\ref{largemass})
for both large  and small $x$.
We then use (\ref{interpolating}) to
numerically perform the sum\footnote{As in section \ref{nonsusy}, for
the $SU(2)$ theory at hand,
     one needs $N > 60$; the linear dependence appears already for
$R/\rho >
{\cal{O}}(10)$.}
over the deconstructed mass spectrum in
(\ref{scalar1})
    and find the following $\rho$-dependence of $\Delta S_{eff}^\prime$:
\beq
\label{scalar5}
\Delta S_{eff}^\prime\big\vert_{interp.} \approx - 0.22 ~ {R \over
      \rho}~,
\eeq
where the coefficient has a slight dependence of $R/\rho$  (becoming
weaker as $R/\rho$ is increased).
We can now compare
this contribution
to the
result found in the step-function approximation with KK spectrum,
i.e.  eqn.~(\ref{seffKK}) adapted to the scalar case by replacing $2
\rightarrow 1/3$:
\beq
\label{scalar4}
\Delta S_{eff}^\prime\big\vert_{step,KK}  =  -  {1\over 3} ~{R \over
\rho}~.
\eeq
We conclude that $\Delta S^\prime_{eff}$ is a linearly
decreasing function of $R/\rho$ even when threshold
corrections are taken into account (albeit with a somewhat
reduced coefficient\footnote{The change in the coefficient
   is mostly due to the better approximation (\ref{interpolating}); the 
result is insensitive
to whether deconstructed or KK spectrum is used, so long as $1/\rho$ is 
sufficiently below
the top of the KK tower.}
  when compared to step-function approximation).

The fact that corrections do not modify our conclusion qualitatively
holds  also for scalars in other representations as well as for fermions
and gauge fields.  Indeed, the propagators for massive
fermions and gauge fields can be related to those of the
scalar fields  \cite{Brown:bt}. In particular, the contribution
of any massive state to the effective action is the same as
that of a scalar up to overall spin- and isospin-dependent
factor (and to the possible would-be zero mode contributions)
\cite{Kwon:2000kf}. Thus, properly including the
contributions of the massive thresholds, we obtain an
improved expression for the one-loop effective action
(\ref{seffgeneral2}):
\beq
\label{seffgeneral3}
      S_{eff}  &=&  {8 \pi^2 \over g^2(1/R)}  +
      \left(  - 7 + {8 \over 3} N_F^a + {2 \over 3} N_F^f  + {1 \over 3}
N_S^a +{1 \over 6} N_S^f \right) \ln {\rho \over R} \\
       &+&  \left( 2  - {8\over 3} N_F^a -{2\over 3} N_F^f + {2\over 3}
N_S^a + {1\over 3} N_S^f \right) \sum\limits_{k = 1}^{max}
      (g(m_k\rho)+\ln m_k\rho)~,\nonumber
\eeq
with $g(m\rho)$ given by the interpolating function
(\ref{interpolating}).

\section{Acknowledgments}

It is a pleasure to thank Csaba Cs\' aki, Josh Erlich, and Yael Shadmi 
for
collaboration at
the early stages of this project and for many useful
discussions. E.P. thanks Nima Arkani-Hamed, Andy Cohen, and Sergei 
Khlebnikov for
discussions and insightful remarks.
Y.S. is also grateful to Markus Luty, Sandip Pakvasa, and
Xerxes Tata for useful discussions.
Y.S. is supported in part  by DOE grant DE-FG03-92-ER-40701.

       \end{document}